\documentclass[final]{IEEEtran}
\usepackage[utf8]{inputenc}
\usepackage{graphicx}
\usepackage{mathtools}
\usepackage[noadjust]{cite}
\usepackage{amssymb}
\usepackage{mathrsfs} 
\ifCLASSOPTIONcompsoc
    \usepackage[caption=false, font=normalsize, labelfont=sf, textfont=sf]{subfig}
\else
    \usepackage[caption=false, font=footnotesize]{subfig}
\fi
\usepackage{graphicx}
\usepackage{booktabs}
\usepackage{tikz}
    \usetikzlibrary{calc}
    \usetikzlibrary{positioning}
    \usetikzlibrary{arrows}
    \usetikzlibrary{decorations.markings}
\usepackage{tikz-3dplot}
    \usetikzlibrary{arrows.meta}
\usepackage{pgfplots}
    \usepgfplotslibrary{groupplots} 
    \pgfplotsset{compat=newest}
\usepackage{siunitx}
\usepackage{bm} 

\usepackage[noabbrev]{cleveref} 

\makeatletter 
\newcommand{\pvec}{\@ifstar{\@pvecstar}{\@pvecnostar}}
\newcommand{\@pvecnostar}[1]{\,\widetilde{\!\bm{#1}}}
\newcommand{\@pvecstar}[2]{{\@pvecnostar{#1}}_{\mkern-1mu\relax#2}}
\makeatother

\title{FDTD Modeling of Periodic Structures: A Review}
\author{Aaron~J.~Kogon, and Costas~D.~Sarris,~\IEEEmembership{Senior Member, IEEE}}

\allowdisplaybreaks[1]
\begin{document}
\maketitle

\begin{abstract}
This paper reviews the state of the art of periodic boundary conditions (PBCs) in Finite-Difference Time-Domain (FDTD) simulations. The mathematical principles and 3D FDTD implementation details are systematically outlined. Techniques for extracting scattering parameters, Brillouin diagrams and attenuation constants are presented, along with the Array Scanning Method (ASM) used to model the interaction of non-periodic sources with periodic structures. Through these techniques, the robustness, utility and efficiency of PBCs are demonstrated and a unified view of the various approaches to the FDTD implementation of PBCs is presented. 
\end{abstract}

\section{Introduction}

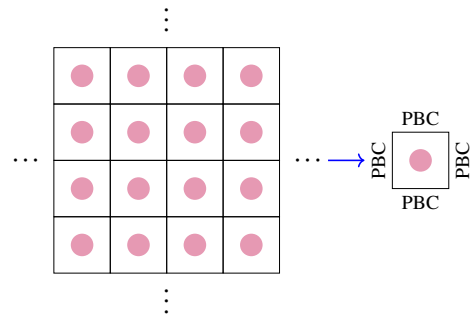
\begin{figure}
    \centering
    \scalebox{0.75}{
    \begin{tikzpicture}
    \foreach \i in {0,...,3}
        {
        \foreach \j in {0,...,3}{
        
        \draw[] (\i,\j) rectangle ++(1,1);
         \fill[white!60!purple] (\i+0.5,\j+0.5) circle (0.2);
        };
        };
        
        \node[] (rightbox) at (4.5, 2) {\(\bm\cdots\)} ;
        \node[] at (-0.5, 2) {\(\bm\cdots\)};
        
        \node[rotate=90] at (2, 4.5) {\(\bm\cdots\)};
        \node[rotate=90] at (2, -0.5) {\(\bm\cdots\)};
        
        \draw[] (6,1.5) -- (6,2.5) node[midway,sloped,above] (cell){PBC} -- (7,2.5) node[midway,sloped,above] (){PBC}--(7,1.5)node[midway,rotate=90,below] (){PBC}--cycle node[midway,sloped,below] (){PBC};
        \fill[white!60!purple] (6.5,2) circle (0.2);
        
        \draw[->,blue,thick] (rightbox) -- (cell);
    \end{tikzpicture}
    }
    \caption{A structure made up of infinitely many identical unit cells (left) can be simulated by a single period bordered by PBCs (right).}
    \label{fig:inf_to_pbc}
\end{figure}

Periodic structures are common in electromagnetics and photonics, appearing in forms such as metamaterials, electromagnetic band gap materials, phased arrays and photonic crystals. Dealing with these structures in the context of the Finite-Difference Time-Domain algorithm (FDTD) is usually handled by simulating a single unit cell terminated by periodic boundary conditions (PBCs) \cite{taflove2005computational}. These techniques collapse an infinitely large periodic domain into one unit cell, thereby prodigiously reducing computational cost of simulation (see \Cref{fig:inf_to_pbc}).

Numerous algorithms exist to simulate periodic behaviour along boundaries. Implementations of periodicity in FDTD are traditionally divided into two groups: direct-field methods, which operate on Maxwell's equations and field-transformation methods, which use auxiliary fields \cite{taflove2005computational}. Many of the periodic boundary condition techniques have serious drawbacks, including the need to modify the standard FDTD algorithm, the need to consume substantial computational resources and the inability to maintain a high Courant stability limit at oblique angles. This paper focuses on PBCs which operate by enforcing constant wavenumbers in periodic directions. We focus on these PBCS since, as explained in \Cref{sec:math_back}, they are broadband, support waves at all angles of incidence and may readily be integrated into the standard FDTD scheme. The PBCs are stable and do not degrade the Courant limit.

PBCs have been used in various applications of interest in electromagnetics and photonics. These include the extraction of band diagrams of periodic negative refractive index metamaterials at microwave frequencies \cite{kokkinos2005periodic}, band diagrams and attenuation constants of leaky wave antennas and leaky coplanar-waveguide slot antennas \cite{kokkinos2006periodic} and band diagrams of many flavours of photonic crystals, including ones with plasmonic modes \cite{luo2003negative, liu2007triangular}. PBCs have also been used to extensively study periodic surface wave antennas \cite{yang2005novel} and antennas over electromagnetic band gap materials (EBGs) \cite{li2010new}. Reflection coefficients of phased arrays of patch antennas were determined in FDTD simulations with PBCs \cite{turner1999fdtd}.

Some confusion has pervaded the literature regarding the nature of PBCs. For example, the sine-cosine method boundary conditions are described as operating at a single frequency and angle \cite{taflove2005computational, harms1994implementation}, but as explained below in \Cref{sec:sinecos}, the sine-cosine algorithm becomes the standard PBC when broadband waves at various angles are excited. The split-field methods in particular has been widely used because of the contention that they are the most natural way in FDTD to simulate general broadband fields interacting with periodic structures at a fixed angle. Broadband fixed-angle simulations of periodic structures can be performed using standard PBCs. This paper systematically approaches FDTD PBCs in detail, and describes principles of their usage. 

PBC algorithms in the literature include the ``constant wavenumber method'' \cite{yang2007simple}, the ``order-\textit{N} method'' \cite{chan1995order}, the ``spatially looped'' algorithm \cite{celuch1995spatially}, ``spectral FDTD'' \cite{aminian2005bandwidth, aminian2006spectral}, and the ``wideband sine-cosine method'' \cite{li2008efficient} among others \cite{turner1999fdtd, lee2001modified, cangellaris1993hybrid, zhou2011efficient}. \Cref{tab:lit_review} tabulates some of the important papers on FDTD PBCs and their applications. Although many of the PBC techniques have been discussed in differing contexts and the precise implementation details of the techniques may vary, they all simulate periodicity through enforcing complex phase shifts between boundaries as described in \Cref{sec:math_back}. This paper is aimed at introducing a general perspective on how PBCs can be translated into FDTD update equations, thereby elucidating the shared foundations of many of the techniques that have been presented over the years. 
\begin{figure*}
    \centering
    \small
    \begin{tabular}{p{0.05\linewidth}p{0.3\linewidth}p{0.5\linewidth}} \toprule
    Paper & Approach & Application \\ \midrule
    \cite{kokkinos2006periodic} & Sine-cosine method & BDs, attenuation constants of metal-strip-loaded dielectric LWAs, leaky coplanar-waveguide-based slot antenna array \\
    \cite{harms1994implementation} & Sine-cosine method & RTS, mode analysis of a single and cross dipole FSSs and a double concentric square loop FSS \\
    \cite{yang2007simple} & Constant wavenumber method & RTS of a dielectric slab and a dipole FSS \\
    \cite{chan1995order} & Order-N method & BDs of a diamond structure PC, effective dielectric constants of various PCs, defect state analysis \\
    \cite{celuch1995spatially} & Spatially-looped algorithm & BDs of corrugated structures \\
    \cite{aminian2006spectral} & Spectral FDTD & RTS of a grounded slab, metallic patches \\
    \cite{li2008efficient} & Sine-cosine method, ASM & Analysis of negative refractive index transmission line perfect lens, microstrip circuits on EBGs \\
    \cite{cangellaris1993hybrid} & Hybrid Spectral/FDTD & BD of corrugated waveguide \\
    \cite{qiang2007asm} & Spectral FDTD, ASM & Analysis of wire EBG fields \\
    \cite{holter1999infinite}  & Pulse-scanning & Active impedance of a dipole array \\
    \cite{roden1998time} & Split-field method & RTS of photonic band gap material \\
    \cite{kao1996finite} & Multiple-grid approach & RTS of square ring and I-slot FSSs \\
    \bottomrule
\end{tabular}

    \caption{Tabulation of some important papers on various types of PBCs and their applications. Abbreviations: ASM: array scanning method,  BD: Brillouin diagram,  FSS: frequency-selective surface, LWA: leaky wave antenna, PC: photonic crystal, RTS: reflection/transmission spectra}
    \label{tab:lit_review}
\end{figure*}

\section{Mathematical Background}\label{sec:math_back}
\subsection{Floquet's Theorem}
The fields interacting with periodic structures are described by Floquet's (or Bloch's) theorem, which dictates that fields in a periodic structure are themselves quasi-periodic, meaning they are periodic up to a particular phase difference and attenuation \cite{ishimaru2017electromagnetic, joannopoulos2008molding}. 

Put mathematically, a field \(\pvec{U}\) propagating along a periodic structure in the \(z\) direction with period \(d_z\) satisfies:
\begin{align}
   \pvec{U}(x, y, z+d_z) = \pvec{U}(x, y, z)e^{-j k_z d_z} \label{eq:floquet_1D}
\end{align}
where the complex value \(k_zd\) describes the change in magnitude and phase between successive periods, and the tilde (\(\ \widetilde{\cdot}\ \)) represents the phasor form of fields.

The function \(\pvec*{U}{p}(x,y,z) = \pvec{U}(x,y,z)e^{j k_zz}\) is periodic (with a period \(d_z\)) and admits a Fourier series expansion
\begin{align}
    \pvec*{U}{p}(x,y,z) = \sum_{n=-\infty}^\infty \bm{A}_{n}(x,y) e^{-2\pi j n z/d_z}
\end{align}
so that:
\begin{align}
    \pvec{U}(x,y,z) = \sum_{n=-\infty}^\infty \bm{A}_{n}(x,y) e^{-j(k_z + 2\pi n/d_z)z}\label{eq:floquet-expansion_1D}
\end{align}
It is seen that $\pvec{U}(x,y,z)$ may be represented as a sum of travelling waves, whose constituent wavenumbers have components $k_{zn} = k_z + 2\pi n/d_z$ in the periodic direction. The constant \(k_{zn}\) therefore corresponds to the \(z\) component of the wavenumber of the \(n^\text{th}\) plane wave. Each mode independently satisfies the Floquet condition (\cref{eq:floquet_1D}). 

Floquet's theorem may likewise describe a field \(\pvec{U}\) propagating in a structure that is periodic in higher dimensions. A general lattice vector of a 3D periodic structure is defined as \(\bm d = n_x d_x \bm{\hat x}+n_y d_y \bm{\hat y}+n_z d_z \bm{\hat z}\) where \(d_x\), \(d_y\) and \(d_z\) are the periods in the \(x\), \(y\) and \(z\) directions, respectively, and \(n_x\), \(n_y\) and  \(n_z\) are integers. Then, in three dimensions \cref{eq:floquet_1D} becomes
\begin{align}
    \pvec{U}(\bm r + \bm d) = \pvec{U}(\bm r)e^{-j \bm k\cdot \bm d} \label{eq:PBC}
\end{align}
where \(\bm k\) is the reciprocal (Bloch) wave vector \(\bm{k} = k_{x}\bm{\hat x} + k_{y}\bm{\hat y} + k_{z}\bm{\hat z}\). As with \cref{eq:floquet-expansion_1D}, the field \(\pvec{U}\) may be expanded into modes:
\begin{align}
    \pvec{U}(\bm r) = \sum_{l,m,n} \bm{A}_{l,m,n} \exp(-j\bm{k}_{l,m,n}\cdot \bm r)\label{eq:floquet-expansion}
\end{align}
where \(\bm{k}_{l,m,n} = k_{xl}\bm{\hat x} + k_{ym}\bm{\hat y} + k_{zn}\bm{\hat z}\) is the reciprocal wave vector.

\subsection{The Problem of FDTD PBC Updates}

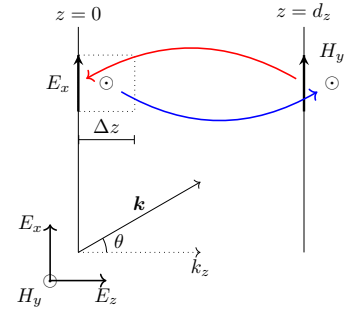
\begin{figure}
    \centering
    \scalebox{0.75}{
    \begin{tikzpicture}
    \tikzset{myptr/.style={decoration={markings,mark=at position 1 with %
    {\arrow[scale=1,>=stealth]{>}}},postaction={decorate}}}
    
        \draw[dotted,->] (0,0) -- (2.5*0.866,0) node[below]{\(k_z\)};

        \draw[->] (0,0) -- (30:2.5) node[above,midway] {\(\bm{k}\)};
        
        \draw[] (0.5,0) arc (0:30:0.5);
        \node at ((15:.75) {\(\theta\)};

        \draw[] (0,0) -- (0,4) node[above] {\(z=0\)};
        
        \draw[] (4,0) -- (4,4) node[above] {\(z=d_z\)};

        \draw[dotted] (0,2.5) rectangle (1, 3.5);
        
        \draw[myptr,very thick] (0,2.5) -- (0,3.5) node[midway](ex_low){} node[left, midway] {\(E_x\)};
        \draw[myptr,very thick] (4,2.5) -- (4,3.5) node[midway](ex_high){};
        
        \draw node[](hy_low) at (.5, 3) {\(\odot\)};
        \draw node[label=above:\(H_y\)] (hy_high) at (4.5, 3) {\(\odot\)};
        
        \draw[red,thick,->] (ex_high) to[bend right] (ex_low) ;
        \draw[blue,thick, ->] (hy_low) to[bend right]  (hy_high);
        
        \draw[|-|] (0, 2) -- (1, 2) node[midway, above] {\(\Delta z\)};

        \draw [->,thick] (-0.5, -0.5) -- (-0.5, 0.5) node[anchor=east] {\(E_x\)};
        \draw [->, thick] (-0.5, -0.5) node{\(\odot\)} node[anchor=north east]{\(H_y\)} -- (0.5, -0.5) node[anchor=north] {\(E_z\)};
    \end{tikzpicture}
    }
    \caption{A unit cell showing two-dimensional FDTD PBC updates (in the \(z\) direction). The \(H_x\) field at \(z=0\) is used to update the corresponding magnetic field at \(z=d_z\) (red arrow). The \(E_y\) field at \(z=d_z+\Delta z/2\) updates the \(E_z\) node at \(z=\Delta z/2\) (blue arrow).}
    \label{fig:updates_1D}
\end{figure}

For simplicity, consider a two-dimensional structure in the \(z\)--\(x\) plane, periodic in the \(z\) dimension, supporting the fields \(E_x\), \(H_y\) and \(E_z\) (see \Cref{fig:updates_1D}). Using \cref{eq:floquet_1D}, the relationships between the two periodic edges \(z=0\) and \(z=d_z\) are
\begin{align}
    \widetilde{E}_x(x, 0) &= \widetilde{E}_x(x, d_z)e^{j k_z d_z} \label{eq:ex_1D}\\
    \widetilde{H}_y(x, d_z + \Delta z/2) &= \widetilde{H}_y(x, \Delta z/2)e^{-j k_z d_z} \label{eq:hy_1D}
\end{align}
in the phasor domain.

As we have seen, the variable \(k_z\) is the component of the wavevector in the \(z\) direction. Whenever \(k_z\) is real, it may be described in terms of the wavenumber \(k = |\bm k|\) by the relationship \(k_z = \omega\cos\theta/v_p\), where \(\theta\) is the angle between the \(z\)-axis and the direction of propagation of the waves and \(v_p\) is the phase velocity of the wave. This relationship allows for \cref{eq:hy_1D,eq:ex_1D} to be inverse Fourier transformed into the time domain:
\begin{align}
    E_x(x, 0, t) &=E_x\left(x,d_z, t+\frac{1}{v_p}d_z\cos\theta\right)\\
    H_y(x, d_z+\Delta z/2, t) &=H_y\left(x,\Delta z/2, t-\frac{1}{v_p}d_z\cos\theta\right).
\end{align}
Whenever \(\cos\theta \neq 0\), one of the update equations requires fields from a future time: \(t+d_z|\cos\theta|/v_p\). This is not feasible in time-domain simulations. The next section discusses how periodicity may be implemented in FDTD by setting a constant wavenumber in the periodic direction.

\subsection{FDTD Implementation}\label{sec:fdtd_imp}

Generally, an electric or magnetic field \(\bm U(\bm r,t)\) may be expressed as the real part of the inverse Fourier transform
\begin{align}
    \bm U_s(\bm r,t) = \frac{1}{2\pi}\int_{-\infty}^\infty \pvec{U}(\bm r,\omega)e^{j\omega t}\,d\omega \label{eq:us_integral}
\end{align}
In a material with a lattice vector \(\bm d\), the field \(\bm U_s\) satisfies the Floquet condition (see \cref{eq:PBC}):
\begin{align}
    \bm U_s(\bm r+\bm d,t) = \bm U_s(\bm r,t)e^{-j\bm k\cdot \bm d}\label{eq:pbc_us}
\end{align}
assuming \(\bm k\) is frequency-independent. Accordingly, field \(\bm U_s\) may also be written as a Floquet series (see \cref{eq:floquet-expansion}):
\begin{align}
    \bm{U}_s(\bm r, t) = \sum_{l,m,n} \bm{A}_{l,m,n}'(t) \exp(-j\bm{k}_{l,m,n}\cdot \bm r)\label{eq:floquet-expansion-td}
\end{align}
where 
\begin{align}
    \bm{A}_{l,m,n}'(t) = \frac{1}{2\pi}\int_{-\infty}^\infty \bm{A}_{l,m,n}(\omega)e^{j\omega t}\,d\omega.
\end{align}

Since the real and imaginary parts of \(\bm U_s\) describe physical fields in quadrature, they may each be simulated using standard FDTD. \Cref{eq:pbc_us} indicates that complex-valued boundary conditions may be used to enforce periodicity on the simulated field \(\bm U_s\) as, for example, in \cref{eq:ex_1D,eq:hy_1D}. Since each mode (in time domain) in \cref{eq:floquet-expansion-td} individually satisfies the Floquet condition, they may all be excited by these boundary conditions.

Translating standard FDTD code into one supporting complex fields \(\bm U_s\) is trivial in programming languages that support complex number arithmetic; the PBCs are complex as described below, but no change to the mesh or update equations is required. Throughout this section, we will refer to the complex-valued fields whose real and imaginary parts correspond to electric and magnetic fields as \(E\) and \(H\).

\begin{figure}
    \centering
    \tdplotsetmaincoords{70}{145}
    \scalebox{0.75}{
    \begin{tikzpicture}[scale=4,tdplot_main_coords,node distance=4cm]
        
    \coordinate (A1) at (0,0,0);
    \coordinate (A2) at (0,1,0);
    \coordinate (A3) at (1,1,0);
    \coordinate (A4) at (1,0,0);
    \coordinate (B1) at (0,0,1);
    \coordinate (B2) at (0,1,1);
    \coordinate (B3) at (1,1,1);
    \coordinate (B4) at (1,0,1);
    
    \coordinate (axis_centre) at (1.1, -0.4, -0.25);
    \coordinate (axis_x) at ($ (axis_centre) + (0.25, 0, 0) $);
    \coordinate (axis_y) at ($ (axis_centre) + (0, 0.25, 0) $);
    \coordinate (axis_z) at ($ (axis_centre) + (0, 0, 0.25) $);

    \draw[->, very thick] (axis_centre) -- (axis_x) node[left]{\(x\)};
    \draw[->, very thick] (axis_centre) -- (axis_y) node[right]{\(y\)};
    \draw[->, very thick] (axis_centre) -- (axis_z) node[above]{\(z\)};
     
    \draw[black,fill=blue!30,opacity=0.8] (A1) -- (A2) -- (B2) -- (B1) -- cycle;
   
    \draw (0, 0.25, 0.5) -- (0, 0.25, 0.75) -- (0, 0.5, 0.75) -- (0, 0.5, 0.5) -- cycle; 
    
    \draw (.25, 0.25, 0.5) -- (.25, 0.25, 0.75) -- (.25, 0.5, 0.75) -- (.25, 0.5, 0.5) -- cycle;
    \draw (0, 0.25, 0.5) -- (0.25, 0.25, 0.5);
    \draw (0, 0.25, 0.75) -- (0.25, 0.25, 0.75);
    \draw (0, 0.5, 0.75) -- (0.25, 0.5, 0.75);
    \draw (0, 0.5, 0.5) -- (0.25, 0.5, 0.5);
    
    \draw[dashed] (0, 0.25, 0) -- (0, 0.25, 1); 
    \draw[dashed] (0, 0.5, 0) -- (0, 0.5, 1);
    \draw[dashed] (0, 0.75, 0) -- (0, 0.75, 1);
    
    \draw[dashed] (0, 0, 0.25) -- (0, 1, 0.25);
    \draw[dashed] (0, 0, 0.5) -- (0, 1, 0.5);
    \draw[dashed] (0, 0, 0.75) -- (0, 1, 0.75);
     
    \draw (A1) -- (A2) -- (B2) -- (B1) -- cycle;
    \draw (.75, 0.25, 0.5) -- (.75, 0.25, 0.75) -- (.75, 0.5, 0.75) -- (.75, 0.5, 0.5) -- cycle;
    \draw (1, 0.25, 0.5) -- (0.75, 0.25, 0.5);
    \draw (1, 0.25, 0.75) -- (0.75, 0.25, 0.75);
    \draw (1, 0.5, 0.75) -- (0.75, 0.5, 0.75);
    \draw (1, 0.5, 0.5) -- (0.75, 0.5, 0.5);
  
    \draw (A3)--(A4)--(B4)--(B3) -- cycle;
    \draw (A1) -- (A4);
    \draw (B1) -- (B4);
    \draw (A2) -- (A3);
    \draw (B2) -- (B3);
        
    \draw[black,fill=red!20,opacity=0.6] (A3) -- (A4) -- (B4) -- (B3) -- cycle;
    
    \draw (1, 0.25, 0.5) -- (1, 0.25, 0.75) -- (1, 0.5, 0.75) -- (1, 0.5, 0.5) -- cycle; 
    
    \draw[dashed] (1, 0.25, 0) -- (1, 0.25, 1); 
    \draw[dashed] (1, 0.5, 0) -- (1, 0.5, 1);
    \draw[dashed] (1, 0.75, 0) -- (1, 0.75, 1);
    
    \draw[dashed] (1, 0, 0.25) -- (1, 1, 0.25);
    \draw[dashed] (1, 0, 0.5) -- (1, 1, 0.5);
    \draw[dashed] (1, 0, 0.75) -- (1, 1, 0.75);

    \end{tikzpicture}
    }
    \caption{The situation of two opposing boundary Yee cells within a unit cell of a periodic structure. The periodic boundaries are shown in colour.}
    \label{fig:fdtd_domain_3d}
\end{figure}
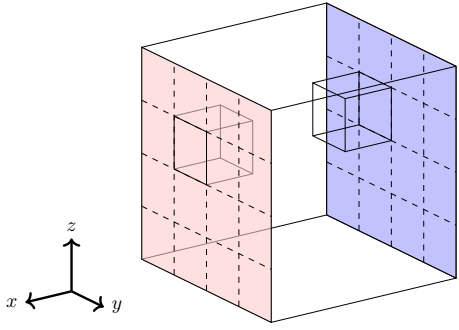

\begin{figure}
    \centering
    \tdplotsetmaincoords{55}{115}
    \scalebox{0.75}{
    \begin{tikzpicture}[scale=4,tdplot_main_coords,node distance=4cm]
    
    \tikzset{e_ptr/.style={-{Latex[scale=1.2,]}, very thick},
    h_ptr/.style={-{Latex[scale=1.2,open]}, very thick},
    }
        
    \coordinate (A1) at (0,0,0);
    \coordinate (A2) at (0,0.5,0);
    \coordinate (A3) at (0.5,0.5,0);
    \coordinate (A4) at (0.5,0,0);
    \coordinate (B1) at (0,0,0.5);
    \coordinate (B2) at (0,0.5,0.5,0.5);
    \coordinate (B3) at (0.5,0.5,0.5);
    \coordinate (B4) at (0.5,0,0.5);
    
    \coordinate (axis_centre) at (1, -0.25, 1);
    \coordinate (axis_x) at ($ (axis_centre) + (0.25, 0, 0) $);
    \coordinate (axis_y) at ($ (axis_centre) + (0, 0.25, 0) $);
    \coordinate (axis_z) at ($ (axis_centre) + (0, 0, 0.25) $);

    \draw[->, very thick] (axis_centre) -- (axis_x) node[left]{\(x\)};
    \draw[->, very thick] (axis_centre) -- (axis_y) node[right]{\(y\)};
    \draw[->, very thick] (axis_centre) -- (axis_z) node[above]{\(z\)};
    
    
    \draw[e_ptr,] ($ (0,0.25,0) - (0,0.1,0)$) -- ($ (0,0.25,0) + (0,0.1,0)$) node[midway](ey_low) {};
    \draw[e_ptr,] ($ (0,0.25,0.5) - (0,0.1,0)$) -- ($ (0,0.25,0.5) + (0,0.1,0)$) node[midway](ey_high) {};
    
    \draw[e_ptr,] ($ (0,0,0.25) - (0,0,0.1)$) -- ($ (0,0,0.25) + (0,0,0.1)$) node[midway](ez_low) {};
    \draw[e_ptr,] ($ (0,0.5,0.25) - (0,0,0.1)$) -- ($ (0,0.5,0.25) + (0,0,0.1)$) node[midway](ez_high) {};
    
    \draw[black,fill=orange!30,opacity=0.8] ($(A1)+(0.25,0,0)$) -- ($(A2)+(0.25,0,0)$) -- ($(B2)+(0.25,0,0)$) -- ($(B1)+(0.25,0,0)$) -- cycle;

    \draw[h_ptr] ($ (0.25,0,0.25) - (0,0.1,0)$) -- ($ (0.25,0,0.25) + (0,0.1,0)$) node[midway](hy_low) {};
    \draw[h_ptr] ($ (0.25,0.5,0.25) - (0,0.1,0)$) -- ($ (0.25,0.5,0.25) + (0,0.1,0)$) node[midway](hy_high) {};
    \draw[h_ptr] ($ ((0.25,0.25,0.5) - (0,0,0.1)$) -- ($ (0.25,0.25,0.5) + (0,0,0.1)$) node[midway](hz_high) {};

    \draw[e_ptr,] ($ (1.5,0.25,0) - (0,0.1,0)$) -- ($ (1.5,0.25,0) + (0,0.1,0)$) node[midway](ey_low_far) {};
    \draw[e_ptr,] ($ (1.5,0.25,0.5) - (0,0.1,0)$) -- ($ (1.5,0.25,0.5) + (0,0.1,0)$) node[midway](ey_high_far) {};
    
    \draw[e_ptr,] ($ (1.5,0,0.25) - (0,0,0.1)$) -- ($ (1.5,0,0.25) + (0,0,0.1)$) node[midway](ez_low_far) {};
    \draw[e_ptr,] ($ (1.5,0.5,0.25) - (0,0,0.1)$) -- ($ (1.5,0.5,0.25) + (0,0,0.1)$) node[midway](ez_high_far) {};
    
    \draw[black,fill=green!20,opacity=0.6]  ($(A3)+(1.25,0,0)$)--($(A4)+(1.25,0,0)$)--($(B4)+(1.25,0,0)$)--($(B3)+(1.25,0,0)$) -- cycle; 

    \draw[h_ptr] ($ (1.75,0,0.25) - (0,0.1,0)$) -- ($ (1.75,0,0.25) + (0,0.1,0)$) node[midway](hy_low_far) {};

    \draw[h_ptr] ($ (1.75,0.25,0) - (0,0,0.1)$) -- ($ (1.75,0.25,0) + (0,0,0.1)$) node[midway](hz_low_far) {};
    
    \draw[h_ptr] ($ (0.25,0.25,0) - (0,0,0.1)$) -- ($ (0.25,0.25,0) + (0,0,0.1)$) node[midway](hz_low) {};
    \draw[white!0!red,thick, ->, shorten >= 0.2cm] (hz_low) to[bend right] (hz_low_far);
    
    \draw[h_ptr] ($ (1.75,0.5,0.25) - (0,0.1,0)$) -- ($ (1.75,0.5,0.25) + (0,0.1,0)$) node[midway](hy_high_far) {};
    
    \draw[h_ptr] ($ (1.75,0.25,0.5) - (0,0,0.1)$) -- ($ (1.75,0.25,0.5) + (0,0,0.1)$) node[midway](hz_high_far) {};

    
    \draw[white!0!red,thick, ->] (hy_low) to[bend right] (hy_low_far);
    
    \draw (A1) -- (A2) -- (B2) -- (B1) -- cycle;
    \draw (A3)--(A4)--(B4)--(B3) -- cycle;
    \draw (A1) -- (A4);
    \draw (B1) -- (B4);
    \draw (A2) -- (A3);
    \draw (B2) -- (B3);
    
    \draw ($(A1)+(1,0,0)$) -- ($(A2)+(1,0,0)$) -- ($(B2)+(1,0,0)$) -- ($(B1)+(1,0,0)$) -- cycle;
    \draw ($(A3)+(1,0,0)$)--($(A4)+(1,0,0)$)--($(B4)+(1,0,0)$)--($(B3)+(1,0,0)$) -- cycle;
    \draw ($(A1)+(1,0,0)$) -- ($(A4)+(1,0,0)$);
    \draw ($(B1)+(1,0,0)$) -- ($(B4)+(1,0,0)$);
    \draw ($(A2)+(1,0,0)$) -- ($(A3)+(1,0,0)$);
    \draw ($(B2)+(1,0,0)$) -- ($(B3)+(1,0,0)$);    
    
    \draw[white!0!red,thick, ->,shorten >= 0.1cm] (hy_high) to[bend left] (hy_high_far);
    \draw[white!0!red,thick, ->, shorten >= 0.2cm] (hz_high) to[bend right] (hz_high_far);
    
        
    \path (.7,0.75,0)
    node[matrix,ampersand replacement=\&,anchor=north west,cells={nodes={font=\sffamily,anchor=west}},
 draw,thick,inner sep=1ex]{
  \draw[e_ptr](0,0) -- ++ (0.6,0); \& \node{\(E\) Node};\\
  \draw[h_ptr](0,0) -- ++ (0.6,0); \& \node{\(H\) Node};\\
  \draw[white!0!red,thick, ->](0,0) -- ++ (0.6,0); \& \node{Update};\\
 };    
    \end{tikzpicture}
    }
    \caption{An illustration of the \(H\) field PBC updates between boundary Yee cells (shown in \Cref{fig:fdtd_domain_3d}). The tangent planes of the \(H\) fields are coloured for clarity.}
    \label{fig:h_updates_3d}
\end{figure}
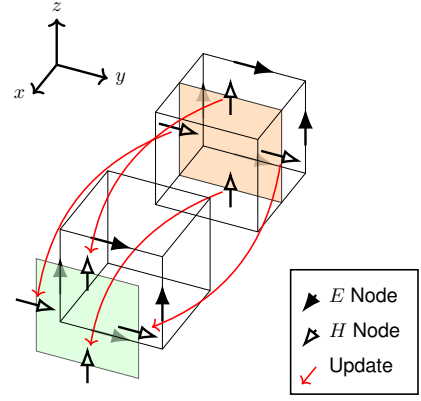

The PBCs may be generalized into three-dimensional update equations. Consider a 3D periodic structure of size \(d_x\times d_y \times d_z\) discretized into \(n_x\times n_y\times n_z\) Yee cells, where each cell is bounded by tangential \(E\) fields and normal \(H\) fields. Let the indices \((i,j,k)\) refer to the physical coordinates \((i\Delta x, j\Delta y, k\Delta z)\).

The \(H\) fields can all be updated from the \(E\) fields, but the boundary \(E\) fields require \(H\) fields from outside the computational domain to be updated. The external auxiliary \(H\)-field nodes can be updated by phase-shifting \(H\) fields from the prior period, which fall within the computational domain. \Cref{fig:fdtd_domain_3d} shows two opposing boundary Yee cells within the computational domain, and \Cref{fig:h_updates_3d} illustrates how the external \(H\) fields are updated between these two Yee cells.

The \(E\) fields residing on the remaining sides (including fields on the boundaries between the updated faces and the non-updated faces) can be computed by directly phase-shifting the respective opposite faces.

Let us choose the faces \(i=n_x+1/2\), \(j=n_y+1/2\) and \(k=n_z+1/2\) to have the external auxiliary \(H\)-field nodes. The auxiliary field nodes are updated as follows:
\begin{align}
    H_x\Bigr|_{i, n_x+1/2, k+1/2}^{n-1/2} &= H_x\Bigr|_{i, 1/2, k+1/2}^{n-1/2}e^{-jk_yd_y}\label{eq:h1}\\
    H_x\Bigr|_{i, j+1/2, n_z+1/2}^{n-1/2} &= H_x\Bigr|_{i, j+1/2, 1/2}^{n-1/2}e^{-jk_z d_z}\label{eq:h2}\\[10pt]
    H_y\Bigr|_{n_z+1/2, j, k+1/2}^{n-1/2} &= H_y\Bigr|_{1/2, j, k+1/2}^{n-1/2}e^{-jk_xd_x}\label{eq:h3}\\
    H_y\Bigr|_{i+1/2, j, n_z+1/2}^{n-1/2} &= H_y\Bigr|_{i+1/2, j, 1/2}^{n-1/2} e^{-jk_zd_z}\label{eq:h4}\\[10pt]
    H_z\Bigr|_{n_x+1/2, j+1/2, k}^{n-1/2} &=H_z\Bigr|_{1/2, j+1/2, k}^{n-1/2}e^{-jk_xd_x}\label{eq:h5}\\
    H_z\Bigr|_{i+1/2, n_y+1/2, k}^{n-1/2}&=H_z\Bigr|_{i+1/2, 1/2, k}^{n-1/2}e^{-jk_yd_y}\label{eq:h6}
\end{align} 

The tangential \(E\) fields on the \(i=0\), \(j=0\) and \(k=0\) planes can subsequently be updated by a direct phase shift:
\begin{align}
    E_x\Bigr|_{i+1/2, 0, k}^n &= E_x\Bigr|_{i+1/2, n_y, k}^n e^{jd_yk_y}\quad (k > 0)\label{eq:e1}\\
    E_x\Bigr|_{i+1/2, j, 0}^n &= E_x\Bigr|_{i+1/2, j, n_z}^n e^{jd_zk_z}\label{eq:e2}\\[10pt]
    E_y\Bigr|_{i, j+1/2, 0}^n &= E_y\Bigr|_{i, j+1/2, 0}^n e^{jd_zk_z}\quad (i > 0)\label{eq:e3}\\
    E_y\Bigr|_{0, j+1/2, k}^n &= E_y\Bigr|_{n_x, j+1/2, k}^n e^{jd_xk_x}\label{eq:e4}\\[10pt]
    E_z\Bigr|_{0, j, k+1/2}^n &= E_z\Bigr|_{n_x, j, k+1/2}^ne^{jd_xk_x}\quad (j > 0)\label{eq:e5}\\
    E_z\Bigr|_{i, 0, k+1/2}^n &= E_z\Bigr|_{i, n_y, k+1/2}^ne^{jd_yk_y}\label{eq:e6}
\end{align}

The \(H\) fields on the faces at \(i=0\), \(j=0\) and \(k=0\) may also be used to update the \(H\) fields on the opposite faces using a phase shift. Alternatively, they may be updated using the standard FDTD updates.

In sum, the FDTD PBC algorithm follows a four-step loop for each time step:
\begin{enumerate}
    \item At time step \(n-1/2\), update all \(H\) fields using standard FDTD, except the auxiliary field nodes at \(i=n_x+1/2\), \(j=n_y+1/2\) and \(k=n_z+1/2\).
    \item Update the auxiliary \(H\)-field nodes by phase-shifting \(H\) fields from \(i=1/2\), \(j=1/2\) and \(k=1/2\).
    \item At time step \(n\), update all \(E\) fields using standard FDTD, except those along \(i=0\), \(j=0\) and \(k=0\).
    \item Update the \(E\) fields along \(i=0\), \(j=0\) and \(k=0\) by phase-shifting the fields on the opposite faces. The phase shift should be the opposite of the phase used in step 2.
\end{enumerate}

In each periodic direction, PBCs set the wavenumber of fields in the periodic direction (up to a multiple of \(2\pi/d\), where \(d\) is the distance between the PBCs). PBCs therefore can be used as boundary conditions to support the propagation of oblique plane waves.

The PBCs can be easily applied to the standard Yee algorithm as a \textit{bona fide} boundary condition. Since there are no modifications to the standard FDTD updates, PBCs do not degrade the Courant limit and can support broadband fields. Since the fields are complex, the computational cost of a PBC simulation is that of two real unit cells. 

Various PBC algorithms exist which simulate complex fields and enforce a constant wavenumber in the periodic directions by using updates as is done above (e.g. \cite{celuch1995spatially, chan1995order, yang2007simple}) or by modifying the FDTD algorithm (e.g. \cite{lee2001modified, cangellaris1993hybrid, aminian2005bandwidth}). All these methods generate the same fields and are in this sense equivalent to each other, though techniques that change the FDTD algorithm suffer from a Courant limit degradation.

Fields anywhere in any period can be accessed by appropriately phase-shifting fields within the primary unit cell by \cref{eq:PBC}. This provides another physical way of understanding PBCs: the fields simulated in the unit cell are those generated by an infinite array of sources, each one progressively phase shifted in accordance with the set PBC wavenumber. Defining \(\bm d = (ld_x, md_y, nd_z)\), an image source in cell \((l,m,n)\) has a phase shift \(e^{-j\bm k \cdot \bm d}\) relative to the ``real'' source in cell \((0,0,0)\) where \(\bm k\) is the wavenumber set by the PBCs.

Hybrid PBC/Perfectly Matched Layer (PML) boundaries have been used to simulate structures with both finite and infinite constituents \cite{li2008efficient, li2009fdtd}. PMLs, which absorb incident waves, can replace PBCs along the regions of the periodic boundary directly facing the finite elements of the periodic structure.

FDTD PBCs hold irrespective of the contents of the unit period. PBCs have been shown to work in skewed domains \cite{yun2000implementation, elmahgoub2010fdtd} and dispersive domains \cite{elmahgoub2012dispersive}. They also have been successfully applied to triangular FDTD meshes \cite{liu2007triangular}.

\subsection{Relationship of Broadband PBCs with the Sine-Cosine Method}\label{sec:sinecos}

Often in the literature, the sine-cosine PBC method \cite{taflove2005computational, harms1994implementation} is described as requiring two simultaneous simulations of the same unit period (call them \(C\) and \(S\)), carrying waves with \(\cos(\omega t)\) and \(\sin(\omega t)\) temporal modulations respectively. At the periodic boundaries, the fields are updated using (see \Cref{fig:updates_1D}):
\begin{align}
    \begin{split}
        \MoveEqLeft E_C(x, y, 0)=\\ \MoveEqLeft[1] \operatorname{Re}\left\{\left(E_C(x, y, d_z)+jE_S(x, y, d_z)\right)e^{jk_zd_z}\right\}
    \end{split}\\
    \begin{split}
        \MoveEqLeft E_S(x, y, 0)=\\ \MoveEqLeft[1] \operatorname{Im}\left\{\left(E_C(x, y, d_z)+jE_S(x, y, d_z)\right)e^{jk_zd_z}\right\}
    \end{split}
\end{align}
\begin{align}
    \begin{split}
        \MoveEqLeft H_C(x, y, d_z+\Delta z/2) =\\
        \MoveEqLeft[1] \operatorname{Re}\left\{\left(H_C(x, y, \Delta z/2)+jH_S(x, y, \Delta z/2)\right)e^{-jk_zd_z}\right\}
    \end{split}\\
    \begin{split}
        \MoveEqLeft H_S(x, y, d_z+\Delta z/2) =\\
        \MoveEqLeft[1] \operatorname{Im}\left\{\left(H_C(x, y, \Delta z/2)+jH_S(x, y, \Delta z/2)\right)e^{-jk_zd_z}\right\}
    \end{split}
\end{align}

Writing \(E_C+jE_S=E\) and \(H_C+jH_S=H\), these update equations correspond precisely to the PBCs presented above (in one dimension). Therefore, sine- and cosine-modulated fields are none other than the physical fields in quadrature described in \Cref{sec:fdtd_imp}. Forcing the real and imaginary fields to have cosine and sine modulations respectively ensures that only the fundamental Floquet mode is excited. However, a broadband excitation is able to excite multiple Floquet modes at once.

\section{Broadband  Reflection and Transmission of Obliquely Incident Plane Waves from Periodic Structures}\label{sec:scattering}
Reflection, transmission and absorption spectra are typically described in terms of angle of incidence and frequency. PBCs are broadband and operate at arbitrary angles, but they are constrained by the relationship \(k=\omega\sin\theta/c\). Except at normal incidence, constant-angled fields are spread out across various frequencies and values of \(k_y\); as the frequency changes, so too must the wavenumber \(k_y\) change to maintain a uniform angle. Likewise, constant-frequency fields are spread out across various angles and \(k_y\) values. Angular and frequency dependence can be decoupled through postprocessing spectra from multiple simulations \cite{aminian2006spectral, yang2007simple}.

To demonstrate the procedure, let us consider a period of a \(y\)-periodic structure, which is extended using periodic boundary conditions along the \(x\)-directed edges and terminated with PMLs along the \(y\)-directed sides (cf. \Cref{fig:grating_schematics}). If an electromagnetic excitation (see below) impinges on the structure, reflected and transmitted fields may then be recorded and transformed into the frequency domain. Thus, to decouple angle and frequency, data from several simulations are required, where \(k_y\) is sampled between \(0\) and \(\omega_\text{max}\sin\theta_\text{max}/c\) (\(\omega_\text{max}\) and \(\theta_\text{max}\) are the maximum angular frequency and maximum angle of incidence, respectively). All these simulations are independent and can be run simultaneously. 


From the simulations, a \(k_y\)--\(f\) diagram of a desired scattering spectrum can be formed. The points on the \(k_y\)--\(f\) diagram intersecting with the curve \(k_y = \omega_0\sin\theta_0/c\) correspond to reflection or transmission spectra at \(\omega=\omega_0\) and \(\theta=\theta_0\) (illustrated in \Cref{fig:kw_grating}). Reflection and transmission under the light line are evanescent in the \(x\) direction and may be ignored.

Scattering parameters from multilayered structures, whose layers may have different periodicities, can be extracted by simulating each layer independently and producing a generalized scattering matrix for each layer. The product of the matrices provides the scattering parameters for the entire system \cite{elmahgoub2011analysis}.
 
\subsection{Excitation}

In a structure with periodic boundary conditions along the \(x\)-directed boundaries, a \(y\)-directed line source excitation of 
\begin{align}
    U(x_0,y) = e^{j(\omega t - k_y y)} \label{eq:plane_source}
\end{align}
produces a plane wave of the form \(e^{j(\omega t - k_xx-k_yy)}\). If the source were only at a point with a \(y\) component at \(l \in (0,d)\), various plane waves would be generated simultaneously; any plane wave whose transverse wavenumber satisfies the equations 
\begin{align}
    k_{y}' = k_y + \frac{2\pi m}{d} = k_y + \frac{2\pi n}{l}
\end{align}
for some \(m,n\in \mathbb{Z}\) would be excited. In other words, a plane wave with a Floquet wavenumber \(k_{yn} = k_y + 2\pi m/d\) would be generated whenever \(ml = nd\) for integers \(m,n\).

A modulated Gaussian pulse is an effective way to produce a broadband excitation \cite{yang2007simple}:
\begin{align}
    S(x_0,y,t) = \exp\left(-\frac{(t-t_0)^2}{t_w^2}+j\omega_\text{avg} t\right)\exp(-jk_yy)
\end{align}
where \(t_w = 2\sqrt{6}/[\pi (f_{\text{max}} - f_{\text{min}})]\) and \(\omega_\text{avg} = 2\pi(f_{\text{max}} + f_{\text{min}})/2\). The time-offset \(t_0\) can be chosen to be 3 or 4 times \(t_w\) so that \(S\) is small at \(t=0\). The maximum frequency should be set to slightly above the frequency of interest so that it is securely within the frequency band. 

\subsection{Numerical Example}

The reflectivity (\(R = |S_{11}|^{2}\)) of an infinite metallic grating was determined at arbitrary angles. The findings were compared to results obtained using other simulation techniques.

To compute reflectivities, simulations were run once with the intervening structure and once without it to enable the calculation of total and scattered fields. Electric and magnetic fields \(\bm E\) and \(\bm H\) were measured along a line \(l\) perpendicular to the PBCs on the same side as the excitation source and each measurement was Fourier transformed into the frequency domain. Finally, power flux across \(l\) was found using the Poynting integral
\begin{align}
    P = \frac{1}{2} \operatorname{Re} \int_S  \left(\pvec{E}(\omega) \times \pvec{H}^*(\omega)\right)\cdot \, d\bm S.
\end{align}

\begin{figure}
    \def\dx{0.06cm}\def\dy{0.06cm} 
    \centering
    \scalebox{0.75}{
    \begin{tikzpicture}[x = \dx, y = \dy] 
        
        \node[draw,anchor=south west,  minimum width=120*\dx, minimum height=32*\dy, label=below:PBC, label=above:PBC] (bound) at (0,0) {};
        
        \filldraw[fill=blue!20,draw=blue] (0,0) rectangle (10, 32) node[pos=.5, rotate=270] {PML};
        
        \filldraw[fill=blue!20,draw=blue] (120,0) rectangle (120-10, 32) node[pos=.5, rotate=270] {PML};
        
        \filldraw[fill=black!20,draw=black] (120-40,8) rectangle (120-41, 24) node[pos=.5, rotate=270,anchor=north] {Grating};

        \draw[line width=2,dotted] (120-65, 0) -- (120-65, 32) node[rotate=270, anchor=north, midway] {Sample};
        
        \draw[line width=2,dotted] (120-80, 0) -- (120-80, 32) node[rotate=270, anchor=north, midway] {Source};
        
        \useasboundingbox (bound.south east) rectangle (bound.north west); 
        
        \draw [->,thick] (-4, -4) -- (-4, 5) node[anchor=east] {\(H_y\)};
        \draw [->, thick] (-4, -4) node{\(\odot\)} node[anchor=north east]{\(E_z\)} -- (5, -4) node[anchor=north west,shift={(-3,0)}] {\(H_x\)};
    \end{tikzpicture}
    }
    \caption{Schematic of the computational domain of the metallic grating.}
    \label{fig:grating_schematics} 
\end{figure}
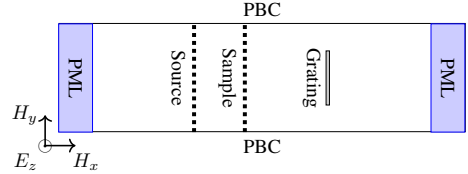

The reflectivity of a periodic metallic grating was determined as another example to illustrate the usage of the constant-wavenumber approach. The simulation took place in a two-dimensional (\(x\)--\(y\)) transverse electric (TE) domain of 33 \(\times\) 120 cells (see \Cref{fig:grating_schematics}). The frequency \(f_\text{max}\) was set to \SI{15}{\GHz}, and the FDTD cell sizes were defined as \(\lambda_{\text{min}}/20\). The time step was determined by a Courant reduction number of \(0.9\). PBCs lined the \(x\)-parallel boundaries and 10-cell-wide PMLs were situated along the \(y\)-parallel boundaries. A perfect electric conductor (PEC) grating of \(1 \times 16\) cells was situated 40 cells from the rightmost boundary, followed 25 cells later by the measurement line, and 15 cells later by the source line. 

The maximum transverse wavenumber \(k_{y,\text{max}}\) was fixed to \(\omega_{\text{max}}/c\) where \(\omega_{\text{max}}\) is the maximum angular frequency of interest, so that the entire range of frequencies would be available for any angled cut. The wavenumber \(k_y\) was sampled uniformly between \(0\) and \(k_{y,\text{max}}\). Each simulation was run for \(2^{12}\) time steps.

\begin{figure}
    \centering
    \def\xpos{1.07}
    \def\ypos{0.77}
    \scalebox{0.75}{
    \begin{tikzpicture}
        \node[anchor=south west,inner sep=0] at (0,0) {        \includegraphics[width=\linewidth]{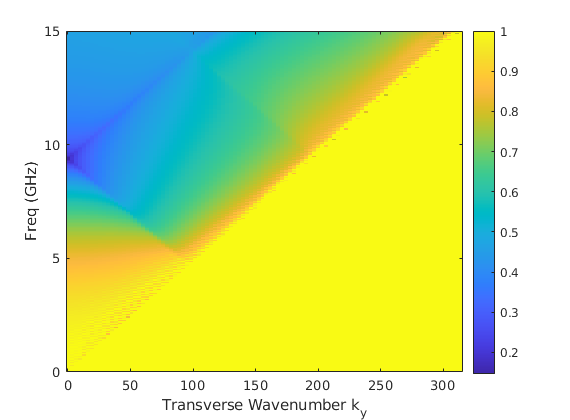}};
        \draw[very thick, white, densely dashed] (\xpos,\ypos) -- (5.5,6.2) node[pos=0.8, above, sloped] {\(\theta = \theta_0\)};
        
        \draw[very thick, black, densely dashed] (\xpos,4) -- (4.7,4) node[pos=0.4, above] {\(\omega=\omega_0\)};

        
    \end{tikzpicture}
    }
     \caption{Computationally derived \(k_y\)--\(f\) diagram of the reflectivity of a metallic grating (\(N_{k_y} = 100\) samples). The reflectivity below the light line was set to unity. The dashed white line illustrates the path along which the reflectivities correspond to a constant angle of incidence \(\theta_0\) over a range of frequencies. The horizontal black line illustrates the path along which the reflectivities correspond to a constant frequency \(\omega_0\) over a range of angles.}
    \label{fig:kw_grating}
\end{figure}

\begin{figure}[t]
  \begin{minipage}{\columnwidth}
    \subfloat[]{
    \begin{tikzpicture}
      \begin{axis}[
          width=\linewidth, 
          height=0.6\linewidth,
          grid=major, 
          grid style={dashed,gray!30},
          xmin=0,xmax=15,
          ylabel={Reflectivity},
          xlabel={Frequency (GHz)},
          x tick label style={rotate=0,anchor=north}
        ]
        \addplot [color=blue] table[x=f,y=r,col sep=comma] {scatter/grate_norm_tfsf.csv}; 
        \addplot [color=orange] table[x=f,y=r,col sep=comma] {scatter/grate_norm_25k.csv}; 
        \legend{Full Size, PBC}
      \end{axis}
    \end{tikzpicture}
    \label{fig:grating_normal}
    }
    \end{minipage}
    \par\bigskip
    \begin{minipage}{\columnwidth}
    \subfloat[]{
    \begin{tikzpicture}
      \begin{axis}[
          width=\linewidth, 
          height=0.6\linewidth,
          grid=major, 
          grid style={dashed,gray!30},
          xmin=0,xmax=15,
          ylabel={Reflectivity},
          xlabel={Frequency (GHz)},
          x tick label style={rotate=0,anchor=north}
        ]
        
        \addplot [color=blue] table[x=f,y=r,col sep=comma] {scatter/grate_pi8_tfsf.csv}; 

        \addplot [color=green] table[x=f,y=r,col sep=comma] {scatter/grate_pi8_25k.csv}; 
        
        \addplot [color=black] table[x=f,y=r,col sep=comma] {scatter/grate_pi8_50k.csv}; 
        
        \addplot [color=orange] table[x=f,y=r,col sep=comma] {scatter/grate_pi8_100k.csv}; 
        \legend{Full Size, PBC (25), PBC (50), PBC (100)}
      \end{axis}
    \end{tikzpicture}
    \label{fig:grating_pi8}
    }
    \end{minipage}
    
    \caption{Reflectivity of a metallic grating at \(\theta = 0\) rad incidence (a) and at \(\theta = \pi/8\) rad incidence (b). The figures compare reflectivity determined by a simulation of a 60 unit cells of the grating (blue) and reflectivity determined by using PBCs. The number of wavenumber samples (\(N_{k_y}\)) is shown in parentheses in (b). (At normal incidence, \(N_{k_y}\) is irrelevant.) }
    \label{fig:grating_refl}
\end{figure}
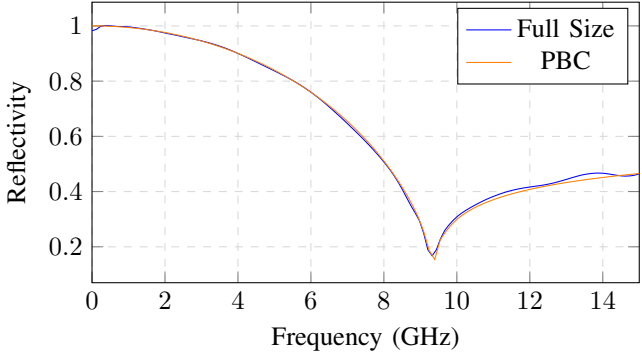
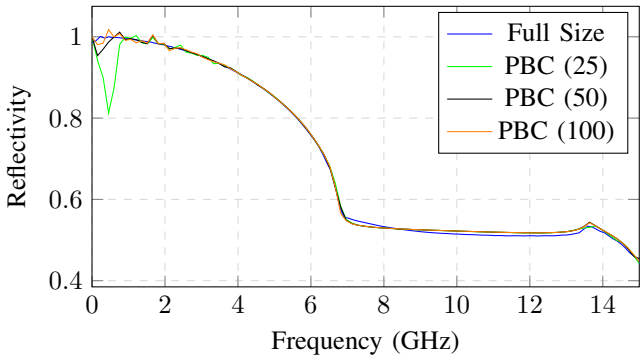

The \(k_y\)--\(f\) diagram appears in \Cref{fig:kw_grating}. The boundary at which \(x\)-directed waves become evanescent (\(\omega = k_y c\)) (the light line) bisects each image. Below the light line, the reflectivity was set to unity. The reflectivities at \(\theta=0\) and \(\theta = \pi/8\) are shown in \Cref{fig:grating_refl} at various \(k_y\) sampling rates. The simulations are confirmed by simulations of 60 periods.

Good agreement exists between  reflectivities of the finite-size simulation and the PBC simulations. There are some artifacts at low frequencies due to proximity of the reflectivity values with the light line. Due to this phenomenon, errors are exacerbated at high angles of incidence. Increasing the \(k_y\) sampling rate diminishes the low-frequency ripples. 

The frequency range in reflectivity plots visibly captures various diffraction regimes. In the normal incidence case, diffraction begins at \SI{9.38}{GHz} corresponding to the dip in the reflectivity diagram. In the \(\pi/8\) rad incident wave case, diffraction begins around the bend at \SI{6.78}{GHz}, and further modes materialize around \SI{13.6}{GHz}, where a knee appears in the plot.

\section{Brillouin (Band) Diagrams}
\label{sec:brill}
In a periodic structure, only certain frequencies are allowed for a particular wavenumber. A diagram depicting the allowed frequencies at a continuum of wavenumbers is called a Brillouin diagram \cite{joannopoulos2008molding}. Brillouin diagrams are often useful for identifying photonic band gaps and dispersive characteristics of materials. 

Since PBCs are broadband and enforce one particular wavenumber, they can be used expeditiously to produce Brillouin diagrams \cite{kokkinos2004rigorous, chan1995order, fan1996large}. When a broadband excitation (spanning a frequency band of interest) in a periodic simulation evolves, some frequencies decay, while others remain, as in a resonant chamber. Extant frequencies correspond to eigenfrequencies on the Brillouin diagram at the wavenumber set by the PBCs. These frequencies appear as peaks in the Fourier transform of a temporal waveform of the electric or magnetic field sampled at a grid point in the unit cell, as illustrated in \Cref{fig:band_diagram}. The sampling should begin only after the transient response of the structure has decayed. Extrapolation techniques, such as the matrix pencil, filter diagonalization, Pad\'e and Prony methods, may be used to calculate the peaks of the Fourier transform from very few time steps \cite{hua1989generalized, mandelshtam1997harmonic, guo2001computation, pereda1992computation}.

One must be sure to place the source excitations and sampling points away from symmetry planes in order to excite and measure fields of various symmetry configurations. To this end, multiple sources and sample points can be used. If a particular symmetry is desired, however, the source can be placed along a symmetry plane accordingly.

\begin{figure}
    \centering
    \begin{tikzpicture}
        \fill[white!60!purple] (0,0) circle (0.4) node[coordinate] (n1){};
        \fill[white!60!purple] (1,0) circle (0.4) node[coordinate] (n2){};

        \fill[white!60!purple] (-0.5,1) circle (0.4) node(n3){};
        \fill[white!60!purple] (0.5,1) circle (0.4) node[coordinate] (n4){};
        \fill[white!60!purple] (1.5,1) circle (0.4) node[coordinate] (n5){};
        
        \fill[white!60!purple] (0,2) circle (0.4) node[coordinate] (n6){};
        \fill[white!60!purple] (1,2) circle (0.4) node[coordinate] (n7){};
        
        \fill[draw=black,white!0!green] (0.75,0.5) circle (0.05) node[coordinate] (s1){};
        \fill[white!0!green] (1.25,1.5) circle (0.05) node[coordinate] (s2){};
        \fill[white!0!green] (0.25,1.5) circle (0.05) node[coordinate] (s3){};
        \fill[white!0!green] (-0.25,0.5) circle (0.05) node[coordinate] (s4){};

        \draw[dashed] (n1) -- (n2) -- (n5) -- (n4) -- cycle node[midway](trap){};
        
        \draw[dashed] (n1) -- (n2) -- (n7) -- (n6) -- cycle node[midway] (centre_rect){};
        
        \def\r{2.5}
        \begin{scope}
            \draw[clip] ($(n1)+(\r,0)$) -- ($(n2)+(\r,0)$) node[midway](cut_trap){} -- ($(n5)+(\r,0)$) -- ($(n4)+(\r,0)$) -- cycle;
            
            \fill[white!60!purple] ($(n1)+(\r,0)$) circle (0.4) node[coordinate] (r1){};
            
            \fill[white!60!purple] ($(n2)+(\r,0)$) circle (0.4) node[coordinate] (r2){};
            
            \fill[white!60!purple] ($(n4)+(\r,0)$) circle (0.4) node[coordinate] (r4){};
            
            \fill[white!60!purple] ($(n5)+(\r,0)$) circle (0.4) node[coordinate] (r5){};
            
            \fill[white!60!green] ($(s1)+(\r,0)$) circle (0.05) node[coordinate] (s_r1){};
        \end{scope}
        
        \def\l{-2.5}
        \begin{scope}
            \draw[clip] ($(n1)+(\l,0)$) -- ($(n2)+(\l,0)$) node[midway](cut_rect){} -- ($(n7)+(\l,0)$) -- ($(n6)+(\l,0)$) -- cycle;
            
            \fill[white!60!purple] ($(n1)+(\l,0)$) circle (0.4) node[coordinate] (l1){};
            
            \fill[white!60!purple] ($(n2)+(\l,0)$) circle (0.4) node[coordinate] (l2){};
            
            \fill[white!60!purple] ($(n4)+(\l,0)$) circle (0.4) node[coordinate] (l4){};
            
            \fill[white!60!purple] ($(n5)+(\l,0)$) circle (0.4) node[coordinate] (l5){};
            
            \fill[white!60!purple] ($(n6)+(\l,0)$) circle (0.4) node[coordinate] (l6){};
            
            \fill[white!60!purple] ($(n7)+(\l,0)$) circle (0.4) node[coordinate] (l7){};
            
            \fill[white!60!green] ($(s1)+(\l,0)$) circle (0.05) node[coordinate] (s_l1){};
            
            \fill[white!60!green] ($(s3)+(\l,0)$) circle (0.05) node[coordinate] (s_l3){};
        \end{scope}
        
        \node[below=0cm of cut_rect]{Supercell};
        \node[below=0cm of cut_trap]{Primitive cell};
    \end{tikzpicture}
    \caption{An illustration of a photonic crystal made up of staggered pink circles (centre) with sources shown in green in each primitive period. The non-orthogonal minimal periodic unit cell is shown at right. A rectangular periodic supercell appears on the left, containing two sources. The boundaries of both the primitive cell and the supercell are shown as dashed lines on the photonic crystal structure.}
    \label{fig:supercell}
\end{figure}
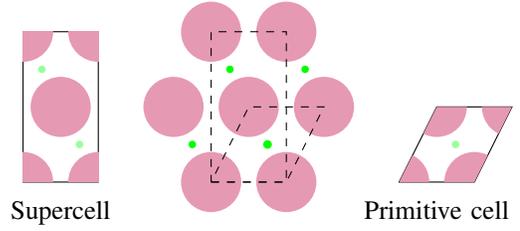

The FDTD algorithm can be modified to accommodate cases where periodic structures are non-orthogonal \cite{taflove2005computational, qiu2000nonorthogonal, ward2000program}. A simpler technique is to enclose a non-orthogonal unit cell in a larger orthogonal ``supercell,'' which too is periodic (see \Cref{fig:supercell}). The supercell technique generates numerous spurious (folded) modes, unless the original source is duplicated (with proper phase modulation) in every primitive period contained in the supercell \cite{taflove2005computational}. Consequently, in \Cref{fig:supercell}, the primitive unit cell contains a single source, while the rectangular supercell contains two. Exponential extrapolation methods have been effectively used to sift away remains of low-amplitude spurious modes. Modes extracted at two points separated by a lattice vector that do not have a \(e^{-j\bm k\cdot \bm d}\) relationship may also be discarded \cite[p. 89]{taflove2013advances}. 

\subsection{Numerical Example}

\begin{figure}
  \begin{center}
  \scalebox{0.9}{
    \begin{tikzpicture}
      \begin{groupplot}[
         group style = {group size = 2 by 1,horizontal sep=0},
         width=\linewidth, 
         height=.8\linewidth,
          ymin=0,ymax=0.9,
          ymajorgrids,
          grid style={dashed,gray!40},
        ]
        \nextgroupplot[ylabel=Frequency \(\omega d/2\pi c\), xmin = 0, xmax = 1, width=0.4\linewidth,xlabel={\(E\) (dB)}, xtick distance=0.5]
        \addplot [red] table[x=e,y=x,col sep=comma] {band/abs_e_X.5.csv}; 
        \addplot [green] table[x=e,y=x,col sep=comma] {band/abs_e_G.5.csv}; 
          
        \nextgroupplot[xmin = 0, xmax = 3, ymax = 0.9,               ylabel=\empty,
            width=0.8\linewidth,
            xtick = {0,1,2,3},
            xticklabels = {\(\Gamma\),X,M,\(\Gamma\)},
            xmajorgrids, yticklabels={,,},
            clip mode=individual,
            xticklabel pos=upper,
            ]
            
        \addplot [only marks, mark = *, color=yellow] table[x=x,y=y,col sep=tab] {band/cylinder_band_book.csv}; 
        \addplot [only marks, mark = +, color=blue] table[x=zone,y=norm_freq,col sep=comma] {band/cylinder_band_fdtd.csv}; 
        
        \draw [ultra thick, draw=green,opacity=0.4] 
        (axis cs: 0.5264, 0) -- (axis cs: 0.5264, 0.9);
        \draw [ultra thick, draw=red,opacity=0.4] 
        (axis cs: 1.47368, 0) -- (axis cs: 1.47368, 0.9);

      \end{groupplot}

      \coordinate (A) at (5.25,0.3);
      \coordinate (B) at (5.25,1.55);
      \coordinate (C) at (4,1.55);
      \coordinate (D) at (4,0.3);
      
      \filldraw[fill=white, draw=black] (A) rectangle (C);
      \draw[black] ($(A)!0.5!(B)$) -- ($(C)!0.5!(D)$);
      \draw[black] ($(B)!0.5!(C)$) -- ($(A)!0.5!(D)$);
      
      \filldraw[fill=blue!20,draw=blue] ($(A)!0.5!(C)$) node[anchor=north east]{\small$\Gamma$} -- (B) node[anchor=south west]{\small X} -- ($(A)!0.5!(B)$) node[anchor=north west]{\small M} -- ($(A)!0.5!(C)$) -- cycle;
    \end{tikzpicture}
    }
    
    \caption{\(\text{TM}_{z}\) Brillouin diagram of cylindrical dielectric columns in a 2D square grid (right). The yellow points are data from \cite{joannopoulos2008molding}; the blue crosses are from the FDTD simulation using PBCs. The inset shows the irreducible Brillouin zone in blue. Note the appearance of a fifth band, not included in data from \cite{joannopoulos2008molding}. The graphs on the left show the Fourier transforms of the electric field samples at two arbitrary points on the irreducible Brillouin zone (normalized by maximum and minimum). The particular points on the Brillouin zone are indicated by the vertical lines of matching colours on the right. The peaks of the fields align with points on the Brillouin diagram intersecting with the vertical lines. Note the small spurious peak on the green plot.}
    \label{fig:band_diagram}
  \end{center}
\end{figure}
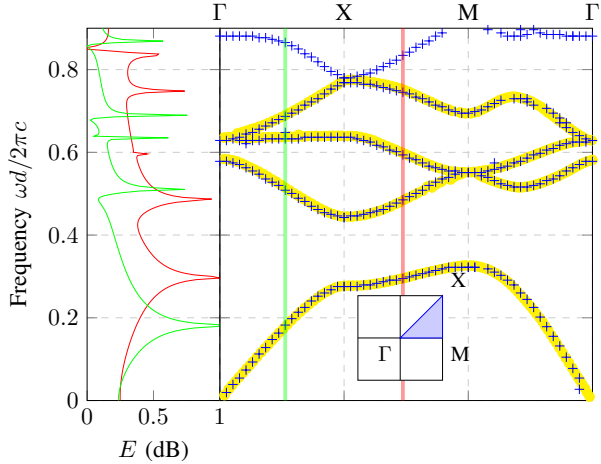

The transverse magnetic (with respect to \(z\); \(\text{TM}_{z}\)) photonic band diagram for a two-dimensional array of infinitely-long dielectric cylinders (\(\varepsilon_r = 8.9\), as alumina) in air was determined using PBCs in FDTD (see \Cref{fig:band_diagram}). The periodicity was \(d = \SI{2}{cm}\) in both the \(x\) and \(y\) planes, and the radius of the cylinder was defined to be \(r = 0.2 d\).

The FDTD cell dimensions were \(\lambda_\text{min}/20\) in the \(x\) and \(y\) axes, and the Courant stability factor was 0.9. The cylinder was placed in the centre of the domain with a staircased boundary. An \(E_z\) source was arbitrarily placed at cell \((17, 10)\) and had a maximum frequency of \SI{15}{GHz} and the sum of samples at two arbitrarily-placed cells was taken after the source decayed. Twenty samples of the wavenumber were taken between each edge of the Brillouin zone, and each simulation was run for \(2^{13}\) time steps.

The Fourier transform technique was used to identify the eigenfrequencies. The frequency peaks of the Fourier transform were determined using MATLAB's \texttt{findpeaks()} function, where peak prominence was set to 0.75.

The band diagram in \Cref{fig:band_diagram} shows good agreement with data from the literature, exhibiting, in particular, a band gap between the first and second modes (31.4\% gap-midgap ratio) \cite{joannopoulos2008molding}. A few spurious frequencies appear sporadically as solitary points in the band diagram. Usually these spurious eigenfrequencies appear as small and smooth peaks in the frequency domain, as shown on the green plot in \Cref{fig:band_diagram}. The spurious points can be removed from the Brillouin diagram by manual deletion, by lengthening simulation time and by calibrating peak detection.

\section{The Array Scanning Method (ASM)}\label{sec:asm}
Let us consider a structure that is periodic in the \(y\) direction, with a period \(d_y\) (for example, see \Cref{fig:grating_schematics}). In this structure, we consider an array of linearly phase-shifted infinite sources at \(\bm{r}_{0} + n d_y\bm y\), where \(r_0\) lies in the \(n=0\) periodic cell and \(n \in\mathbb Z\). Then, the array of sources satisfies the periodic boundary conditions at a particular \(k_y\). An electric or magnetic field \(\pvec U_\infty(x, y, t, k_y)\) generated by the array can be described as a superposition of the fields generated in each period:
\begin{align}
    \bm U_\infty(x,y,t,k_y) = \sum_{n=-\infty}^\infty \bm {U}_{n}(x,y,t) e^{-jk_ynd_y}\label{eq:asm_super}
\end{align}
Here, the field due to the source in cell \(n\) is \(\bm {U}_{n}(x,y,t) = \bm {U}_{0}(x,y-nd_y,t)\) where \(\bm {U}_{0}\) is the field generated by a source in the \(n=0\) cell.

The field \(\bm {U}_{n}\) due to sources in cell \(n\) alone can be extracted by orthogonality of \(e^{-jk_ymd_y}\) over the Brillouin zone:
\begin{align}
    \bm {U}_{n}(x, y, t) = \frac{d_y}{2\pi}\int_{-\pi/d_y}^{\pi/d_y} \bm {U}_{\infty}(x, y, t, k_y) e^{jk_y n d_y} \,dk_y\label{eq:asm}
\end{align}
The integral can be generalized to account for periodicity in other axes by integrating along the Brillouin zones in other periodic directions. Defining \(\bm{d} = (ld_x, md_y, nd_z)\) the integral
\begin{align}
\begin{split}
    \bm {U}_{l,m,n}(\bm{r}, t) = &\frac{d_x d_y d_z}{(2\pi)^3} \int_{-\pi/d_x}^{\pi/d_x} \int_{-\pi/d_y}^{\pi/d_y} \int_{-\pi/d_z}^{\pi/d_z} \\
    &\bm {U}_{\infty}(\bm{r}, t, \bm{k}) e^{j \bm{k} \cdot \bm{d}} \,d^3\bm{k}\label{eq:asm_3d}
\end{split}
\end{align}
determines the field \(\bm{U}_{l,m,n}\) due to a source in the \((l,m,n)^\text{th}\) cell.

This technique is known as the Array Scanning Method (ASM) \cite{munk1979plane}. Note that a sum of integrals as in \cref{eq:asm,eq:asm_3d} can be used to compute fields generated by an arbitrary number of sources in various periods. 

FDTD PBCs explicitly define each wavenumber component of \(\bm{k}\). Consequently, the field \(\bm {U}_{\infty}(\bm{r}, t, \bm{k})\) can be extracted from a number of simulations. The integral in \cref{eq:asm} can then be estimated using a numerical quadrature technique \cite{qiang2007asm, li2008efficient, yang2007simple}. The field \(\pvec*{U}{n}\) obtained through the ASM will be real-valued if the excitation is real-valued as well.

Capolino et al. describe a physical interpretation of numerical quadrature error of the ASM \cite{capolino2007comparison}. They notice that when \(M\) samples of \(k_y\) are used to estimate \cref{eq:asm} due to a source at \(\bm{r}_{0}\) using the midpoint rectangular rule of integration, the resultant field is one generated by sources located periodically at points \(\bm{r}_{0} + nMd_y\bm{\hat y}\). Thus, \(M\) should be chosen so that the periodic image sources are sufficiently far. In \cite{capolino2007comparison}, the midpoint rule is found to perform at least as well as higher-order Gaussian quadrature.

\begin{figure}
  \begin{center}
  \scalebox{0.75}{
    \begin{tikzpicture}
    \tikzset{
      pics/carc/.style args={#1:#2:#3}{
        code={
          \draw[pic actions] (#1:#3) arc(#1:#2:#3);
        }
      }
    }
    
    \draw (0,0) -- (1,0) -- (1,2) -- (0,2) -- cycle;
    \draw (1,0) -- (2,0) -- (2,2) -- (1,2) -- cycle;
    
    \node[] at (2.5,1) {\(\cdots\)};
    \draw (4,0) -- (5,0) -- (5,2) -- (4,2) -- cycle;
    \draw (3,0) -- (4,0) -- (4,2) -- (3,2) -- cycle;
    
    \filldraw[black] (0.75,1) circle (2pt) node[rotate=270, anchor=north] {Source};
    
    \filldraw[black] (4.75,1) circle (2pt) node[rotate=270, anchor=north] {Image};
    
    \draw[thick, dashed] (1.25, 0) -- (1.25, 2);
    \draw[thick, dashed] (0.25, 0) -- (0.25, 2);
    
    \draw[|-|] (0.75,2.15) -- (1.25,2.15) node[anchor=south,midway]{\(a\)};
    
    \draw[|-|] (0, -0.15) -- (1, -0.15) node[anchor=north,midway]{\(d\)};
    
    \draw[|-|] (1.25, -0.15) -- (4.75, -0.15) node[anchor=north,midway]{\(Md-a\)};
    
    \draw[thick, red] (0.75,1) pic{carc=-15:15:3.2};
    \draw[thick, red] (0.75,1) pic{carc=-15:15:3.3};
    \draw[thick, red] (0.75,1) pic{carc=-15:15:3.4};
    
    \draw[thick, red] (4.75,1) pic{carc=165:195:3.2};
    \draw[thick, red] (4.75,1) pic{carc=165:195:3.3};
    \draw[thick, red] (4.75,1) pic{carc=165:195:3.4};
    
    \end{tikzpicture}
    }
  \end{center}
  \caption{Schematic illustrating how to select ASM order to inhibit parasitic fields from entering a domain of interest. Each box represents a unit cell. The domain of interest is denoted by dashed lines at a distance \(a\) to the left and right of the source. The ASM order \(M\) is high enough such that within the runtime, waves (shown in red) generated by the image source do not enter the domain of interest.}
  \label{fig:ASM_schematic}
\end{figure}
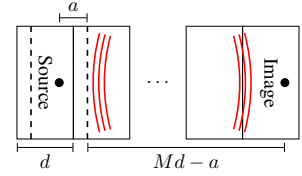

The number of required ASM simulations can be bounded strictly, by ensuring that the initial wavefronts of parasitic image sources cannot enter a region of interest (which can be larger than the actual computational domain) within the total simulation time \(t=t_0\). For a point source turning on at time \(t=0\), consider the distance to the farthest boundary of the region of interest, \(a\) (see \Cref{fig:ASM_schematic}). A parasitic copy of the source will arrive at the boundary of the region of interest at time  \(t = (Md-a)/c\) or later, where \(M\) is the ASM order and \(d\) is the period length. Therefore, \(M\) may conservatively be chosen to be \(M = \lceil (t_0c+a)/d \rceil\), where \(\lceil x \rceil\) equals the least integer greater than or equal to \(x\). 

Even and odd mirror symmetry across a central axis can be exploited to reduce the number of simulations required by a factor of two. Suppose, for instance, that a field generated by a source in cell \(n=0\) is even in the periodic direction \(y\): \(\bm {U}_{0}(x,-y,t) = \bm {U}_{0}(x,y,t)\). The consequential symmetry relation of \(\bm U_n\) becomes \(\bm U_n(x,-y,t) = U_{-n}(x,y,t)\). \Cref{eq:asm_super} implies
\begin{align}
    \bm {U}_{\infty}(x,-y,t,-k_y) &= \sum_{n=-\infty}^\infty \bm {U}_{-n}(x,y,t) e^{jk_ynd_y}\\
    &= \sum_{n=-\infty}^\infty \bm {U}_{n}(x,y,t) e^{-jk_ynd_y}\\
    &= \bm {U}_{\infty}(x,y,t,k_y).
\end{align}
If instead \(\pvec*{U}{0}\) were odd in the \(y\) direction, we would have \(\bm {U}_{\infty}(x,y,t,k_y) = -\bm {U}_{\infty}(x,-y,t,-k_y)\). 

Assuming even symmetry, the integral in \cref{eq:asm} may be broken up at at \(k_y=0\):
\begin{align}
\begin{split}
    \bm {U}_{n}(x, y, t) = 
    \frac{d_y}{2\pi}&\int_{0}^{\pi/d_y} \left( \bm {U}_{\infty}(x, y, t, k_y) e^{jk_y n d_y} + \right.\\
    &\left. \bm {U}_{\infty}(x, -y, t, k_y) e^{-jk_y n d_y}\right)\,dk_y .\label{eq:asm_half_domain}
\end{split}
\end{align}
The sign of the second term in \cref{eq:asm_half_domain} should be negated if \(\bm {U}_{0}\) is odd. In \cref{eq:asm_half_domain}, the domain of integration has been halved, thereby reducing the number of required simulations commensurately.

PMLs terminating periodic structures are known to cause reflections, even with infinite numerical resolution. Although they reflect more than PMLs when situated in free space, matched adiabatic absorbers perform better than PMLs when truncating periodic structures \cite{oskooi2008failure}. PBCs have been explored as absorbing boundary conditions for periodic structures when the ASM is applied. PBCs have been found to have comparable performance to PMLs when used as terminal boundary conditions of a periodic structure \cite{li2009fdtd}. 

\subsection{Numerical Example}

\begin{figure}
  \centering
      \begin{minipage}{\columnwidth}
      \centering
        \subfloat[]{
        \scalebox{0.75}{
        \begin{tikzpicture}
          \begin{axis}[
              width=\textwidth, 
              height=0.8\linewidth,
              grid=major, 
              grid style={dashed,gray!30},
              xmin=-335,xmax=335,
              ymin=-35,
              ylabel={Electric Field Amplitude (dB)},
              xlabel={Transverse (\(y\)) distance (mm)},
              x tick label style={rotate=0,anchor=north},
              ylabel style={alias=ylab},
              yticklabel style={text width=2em,align=right},
            ]
            \addplot [color=blue, ] table[x=x,y=e,col sep=comma] {asm/asm_mag.csv}; 
            \addplot [color=green, ] table[x=x,y=e,col sep=comma] {asm/full_mag_15.csv};
            \addplot [color=black, ] table[x=x,y=e,col sep=comma] {asm/full_mag_20.csv};
            \addplot [color=orange, ] table[x=x,y=e,col sep=comma] {asm/full_mag_26.csv};
            \legend{ASM,15 Periods,20 Periods,26 Periods}
          \end{axis}
    \end{tikzpicture}
    }
    }
    \end{minipage}
    \par\bigskip
    \begin{minipage}{\columnwidth}
        \centering
        \subfloat[]{\def\dx{0.033335cm}\def\dy{0.033335cm}
        \scalebox{0.75}{
        \begin{tikzpicture}[x = \dx, y = \dy]
          \begin{axis}[
              at={(-\textwidth/2+101/2+10, -0.8\linewidth/2-12)},
              width=\textwidth, 
              height=0.8\linewidth,
              grid=major, 
              grid style={dashed,gray!30},
              xmin=-335,xmax=335,
              ylabel={Electric Field Phase (rad)},
              xlabel={Transverse (\(y\)) distance (mm)},
              x tick label style={rotate=0,anchor=north},
              legend pos=south east,
              ylabel style={alias=ylab},
              yticklabel style={text width=2em,align=right},
            ]
            \addplot [color=blue, ] table[x=x,y=phi,col sep=comma] {asm/asm_phase.csv}; 
            \addplot [color=green, ] table[x=x,y=phi,col sep=comma] {asm/full_phase_15.csv};
            \addplot [color=black, ] table[x=x,y=phi,col sep=comma] {asm/full_phase_20.csv};
            \addplot [color=orange, ] table[x=x,y=phi,col sep=comma] {asm/full_phase_26.csv};
            \legend{ASM,15 Periods,20 Periods,26 Periods}
          \end{axis}


         \node[draw,anchor=south west,  minimum width=101*\dx, minimum height=32*\dy, label=below:\small PBC, label=above:\small PBC] (bound) at (0,0) {};
        
        \filldraw[fill=blue!20,draw=blue] (0,0) rectangle (10, 32) node[pos=.5, rotate=270] {\small PML};
        
        \filldraw[fill=blue!20,draw=blue] (101,0) rectangle (101-10, 32) node[pos=.5, rotate=270] {\small PML};
        
        \filldraw[fill=black!20,draw=black] (50,0) rectangle (51, 8);
        
        \filldraw[fill=black!20,draw=black] (101-50,32) rectangle (101-51, 32-8);
        
        \node[rotate=270,anchor=north] at (101-50, 16) {\small Grating};

        \draw[line width=2,dotted] (101-30, 0) -- (101-30, 32) node[rotate=270, anchor=north, midway] {\small Sample};
        
        
        \filldraw[black] (101-70,16) circle (2pt) node[rotate=270, anchor=north] {\small Source};
        
        
        \draw [->,thick] (-8, -8) -- (-8, 3) node[anchor=east] {\(H_y\)};
        \draw [->, thick] (-8, -8) node{\(\odot\)} node[anchor=north east]{\(E_z\)} -- (3, -8) node[anchor=north west,shift={(-3,0)}] {\(H_x\)};
        \end{tikzpicture}
        }
        \label{fig:asm-phase-inset}
    }
    \end{minipage}
    \caption{Electric field amplitude (a) and phase (b) one wavelength after a metallic grating, using the ASM (blue), simulation of 15 periods (green), 20 periods (grey) and with 26 periods (orange). The grating schematic is shown as an inset in (b).}
    \label{fig:asm_fields}
\end{figure}
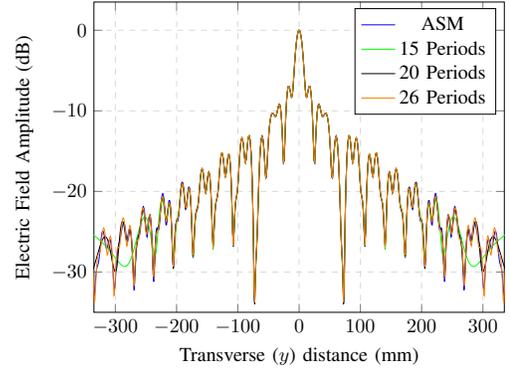
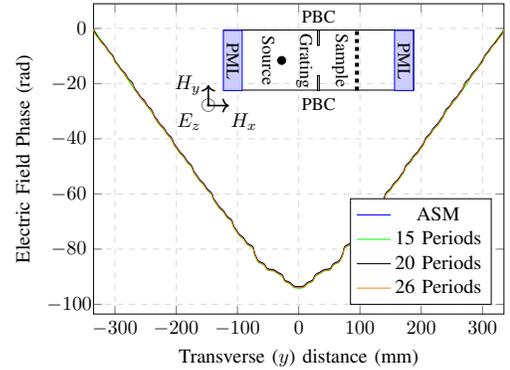

As an example of an application of the ASM, the TE-field transmitted through a metallic grating was calculated. The simulation was run with a source at \SI{15}{GHz} for \(2^{11}\) time steps. The FDTD cell size was set to \(\lambda/20\). The grating comprised of strips \(\SI{1}{mm} \times \SI{16}{mm}\) PEC. One \SI{32}{mm} period of the grating was simulated using PBCs and 10-cell PMLs to extend the domain (shown in the inset of \Cref{fig:asm-phase-inset}). The source was situated along the grating aperture one wavelength to the left of the grating, and frequency domain measurements were taken one wavelength to the right of the grating. The left PML was situated one wavelength to the left of the source point and the right PML was situated one wavelength to the right of the sampling line.

The ASM was used to calculate the fields along the sample line over 21 periods.  Due to the symmetry of the problem, \(k_y\) was varied uniformly 20 times between \(-\pi/d_y\) and \(0\) and integration was carried out using the midpoint rectangular rule. Finite simulations were used to confirm the results of PBC simulations (see \Cref{fig:asm_fields}).

\section{Modelling Attenuating Fields in Periodic Structures}
\label{sec:lossy}
At first sight, PBCs are not well conducive to estimating spatial field decay (as a consequence of loss or leakage), since PBCs explicitly enforce the wavenumber across a period, while the real part of this wavenumber describes the desired loss factor. In other words, the phase and attenuation boundary conditions cannot simultaneously be set. Nonetheless, two solutions have been proposed.

To illustrate the first technique, consider a structure periodic in the \(y\) direction with a period \(d_y\). A simulation of two adjacent periods may then be performed, with PBCs in the periodic direction \cite{kokkinos2006periodic}. If fields \(U\) may be assumed to have a spatial profile of \(e^{-j k_y y} = e^{-\alpha y - j \beta y}\), the wavenumber \(k_y\) may be found from sampling fields one period apart:
\begin{align}
\begin{split}
    k_y(\omega) &= \beta(\omega)-j\alpha(\omega)\\
    &= \frac{j}{d_y}\log\left(\frac{\mathscr{F}[U(x,y+d_y,t)] }{\mathscr{F}[U(x,y,t)]}\right)
    \label{eq:kokkinos_formula}
\end{split}
\end{align}
where \(\mathscr{F}[\cdot]\) denotes a temporal Fourier transform. The real part of \cref{eq:kokkinos_formula} reproduces the phase shift which was set by the PBCs. The real part therefore supplies only redundant information, which may be used to help validate the implementation of the algorithm. However, the imaginary part supplies the attenuation constant.

Another technique uses tools for determining loss in microwave cavities to extract loss of guided waves \cite{xu2007finite}. A \(y\)-periodic guided electric or magnetic field mode takes on the form
\begin{align}
    \bm U = \bm{U}_{0} (\bm r)e^{j(\omega t - k_y y)}
\end{align}
where \(k_y=\beta-j\alpha\). Viewing the periodic unit cell as a lossy cavity, it is apparent that at a given point, fields decay temporally as \(\exp(-\alpha_t t)\) for a particular \(\alpha_t\). If \(t_p\) is the time taken for the envelope of a wave to travel a single period \(d_y\), the temporal and spatial attenuation constants may be related through
\begin{align}
    e^{- \alpha d_y} = e^{- \alpha_t t_p}
\end{align}
so that \(\alpha_t =\alpha d_y/t_p = \alpha v_g\). With this relationship, the field \(\bm U\) may be written as:
\begin{align}
    \bm U = \bm{U}_{0} (\bm r)e^{j(\omega+j\alpha v_g)t - j\beta y}
\end{align}

In this form, the spatial decay constant \(\alpha\) can instead be described in terms of a complex frequency \(\omega' = \omega+j\alpha v_g\). 

The complex frequency \(\omega'\) can be extracted directly by fitting complex exponentials to a temporal sample of a field component using an extrapolation algorithm (see \Cref{sec:brill}). Alternatively, the unit cell simulation can be again seen as a resonant cavity problem, so that the quality factor \(Q\) may be calculated. The full-width half-maximum (FWHM) of the magnitude of the Fourier transform of the sample at \(\omega\) (\(\Delta \omega\)) can be found once the fields have sufficiently decayed. From this, we compute
\begin{align}
    Q = \frac{\omega}{\Delta \omega}.
\end{align}
Then, the relationship
\begin{align}
    Q = \frac{\operatorname{Re}(\omega')}{2\operatorname{Im}(\omega')} = \frac{\omega}{2\alpha v_g}
\end{align}
produces the formula
\begin{align}
    \alpha = \frac{\Delta\omega}{2v_g}.
\end{align}

The group velocity \(v_g\) can be estimated by differentiating the Brillouin diagram (see \Cref{sec:brill}) at \(\beta\). The group velocity can also be estimated by \(v_g\approx\omega/\beta\) when dealing with quasi-TEM modes.

\subsection{Numerical Example}
\begin{figure}
    \def\dx{0.1cm}\def\dy{0.1cm} 
    \centering
    \scalebox{0.75}{
    \begin{tikzpicture}[x = \dx, y = \dy] 
        
        \node[draw,anchor=south west,  minimum width=51*\dx, minimum height=24*\dy, label=below:PBC, label={[anchor=north,rotate=-90]left:PEC}, label=above:PBC, label={[anchor=south,rotate=-90]right:PEC}] (bound) at (0,0) {};

        
        
        
        \filldraw[fill=red!20,draw=black] (0,3) rectangle (51, 14);
        

        \draw[line width=2,dotted] (0,16) -- (51,16) node[rotate=0, anchor=south, midway] {Source};
        
        
        \filldraw[black] (13, 5) circle (2pt) node[rotate=0, anchor=south] {Sample};
        
         \useasboundingbox (bound.south east) rectangle (bound.north west); 
        
        \draw [->,thick] (-4, -4) -- (-4, 5) node[anchor=east] {\(H_y\)};
        \draw [->, thick] (-4, -4) node{\(\odot\)} node[anchor=north east]{\(E_z\)} -- (5, -4) node[anchor=north west,shift={(-3,0)}] {\(H_x\)};
    \end{tikzpicture}
    }
    \vspace{10pt}\caption{Schematic of a period of a waveguide loaded with lossy slabs, which was simulated to illustrate the extraction of attenuation constants. The lossy slab appears in pink.}
    \label{fig:bragg_lossy} 
\end{figure}
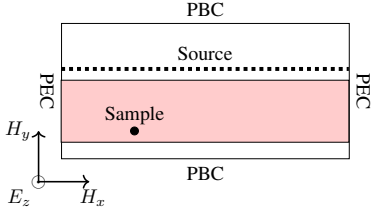

A 2D simulation of a TE\(_1\) mode in a lossy Bragg reflector was carried out in \(\SI{1.2}{mm} \times \SI{3.8}{mm}\) region (Courant factor of 0.9). The periodicity was set to be along the sides parallel to the \(x\) axis, and the sides parallel to the \(y\) axis were set to PEC. The slab thickness was set to \SI{1.6}{mm}, with a relative permittivity of \(\varepsilon_r = 3\) and a conductivity of \(\sigma = \SI{0.05}{\Omega m}\). Each FDTD cell had side lengths of \(\lambda_{\text{min}}/50\). The region was excited by a source, randomly placed at \(y = \SI{16}{cells}\), with a TE\(_1\) spatial profile and a Gaussian temporal profile which would inject frequencies up to \SI{30}{GHz}. A schematic of the unit cell is shown in \Cref{fig:bragg_lossy}.

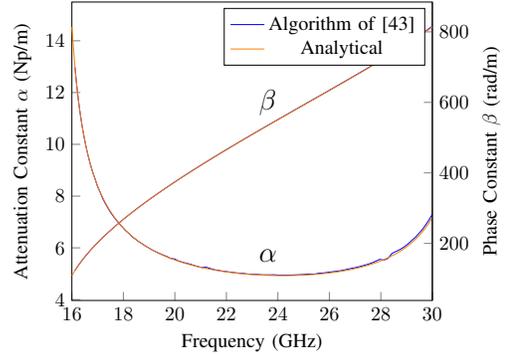
\begin{figure}
  \begin{center}
  \scalebox{0.75}{
    \begin{tikzpicture}
      \begin{axis}[
          width=\linewidth, 
          width=0.9\linewidth,
          xmin=16,xmax=30,
          ylabel={Attenuation Constant \(\alpha\) (Np/m)},
          xlabel={Frequency (GHz)},
          tick pos=left,
        ]
        \addplot [color=blue] table[x=f,y=a,col sep=comma] {lossy/wu_comp_a.csv}; 
        \addplot [color=orange] table[x=f,y=a,col sep=comma] {lossy/wu_analy_a.csv}; 
        
        \node[anchor=west] (source) at (axis cs:23,5.7){\Large\(\alpha\)};
      \end{axis}
      
      \begin{axis}[
          width=\linewidth, 
          width=0.9\linewidth,
          xmin=16,xmax=30,
          hide x axis,
          axis y line*=right,
          ylabel={Phase Constant \(\beta\) (rad/m)},
          xlabel={Frequency (GHz)},
          x tick label style={rotate=0,anchor=north}
          x tick label style={rotate=0,anchor=north}
        ]
        \addplot [color=blue] table[x=f,y=b,col sep=comma] {lossy/wu_comp_b.csv}; 
        \addplot [color=orange] table[x=f,y=b,col sep=comma] {lossy/wu_analy_b.csv}; 
        
        \legend{Algorithm of \cite{xu2007finite}, Analytical}
        
        \node[anchor=west] (source) at (axis cs:23,600){\Large\(\beta\)};
      \end{axis}
    \end{tikzpicture}
    }
    \caption{Attenuation and phase constants of a waveguide loaded with lossy slabs, determined computationally using the algorithm of \cite{xu2007finite} (blue) and by an analytical formula (orange). }
    \label{fig:atten_plot}
  \end{center}
\end{figure}

100 simulations of the unit cell were run, with PBC wavenumbers uniformly selected between \SI{116}{rad/m} and \SI{825}{rad/m}. A sample randomly placed at cells \((15, 6)\) recorded fields for 1000 time steps after the source decayed, and the matrix pencil algorithm was used to determine the complex frequency of the sample \(\omega'\). The group velocity was approximated by differentiating the Brillouin diagram using the central difference derivative. The computed attenuation constant is compared to the analytically derived attenuation constant in \Cref{fig:atten_plot}. Here, the analytical attenuation constant was calculated using the transmission matrix approach \cite{orfanidis2002electromagnetic}.

\section{Summary and Conclusion}

This paper presented, reviewed and synthesized advances in periodic boundary conditions in the FDTD algorithm. 

PBCs operate by enforcing particular wavenumbers in the periodic directions using a complex phase shift. The simulated fields are phasors with temporal and spatial modulation. PBCs are stable and do not affect the Courant limit. The background and derivations propounded above provide a unified basis for many PBC techniques from the literature.

Techniques for extracting scattering parameters and band diagrams were given and illustrated. The array scanning method was outlined as a route to introduce finite sources into an infinitely periodic structure, by running several disjoint simulations of the unit cell. The paper finally described algorithms used to extract the attenuation constants of decaying waves in periodic structures. 

\bibliographystyle{IEEEtran}
\bibliography{bib}

\end{document}